# Optimal Decentralized Economical-sharing Scheme in Islanded AC Microgrids with Cascaded Inverters

Lang Li[1], Huawen Ye[1], Yao Sun[1*], Zhangjie Liu[1], Hua Han[1], Mei Su[1], Josep M. Guerrero[2]

[1] School of Information Science and Engineering, Central South University, Changsha 410083, China
[2] Department of Energy Technology, Aalborg University, DK-9220 Aalborg East, Denmark
* Correspondence: yaosuncsu@gmail.com (Y.S.)

**Abstract:** To address the economical dispatch problem without communications in islanded AC microgrids consisting of cascaded inverters, this paper proposes an optimal decentralized economical-sharing scheme. In proposed scheme, optimal sharing function of the current is applied to generate the reference voltages. And the frequency is used to drive all distributed generators (DGs) synchronize operation in microgrids. When the microgrid is in steady state, DGs share a single common frequency and current in terms of the proposed scheme. Thus the potential advantages of simplicity and decentralized manner are retained. The AC microgrid model has been developed through simulations and experiments to verify the effectiveness and performance of the proposed scheme.

## 1. Introduction

Recently, interests have been concentrated on integrating distributed generators due to many economical challenges, technological advancements and environmental impacts [1-4]. The microgrid technology that integrates DGs, energy storage elements and loads has become a most effective way to solve the permeation of large-scale DGs to power grid [5-6]. Usually, there are different types of DGs in a microgrid, and their generation costs are various [7-8]. From the perspective of economics, less costly DGs should be controlled to provide more power and all DGs should be coordinated in economical operation modes.

Economical operation schemes for microgrid could be classified into the centralized, distributed and decentralized approaches. The centralized schemes hold the advantages of economy, better voltage quality and frequency regulations in microgrids [9-11]. However, the control decisions depend on complicated central controllers and communications, which increase costs and complexity, and reduce reliability [10].

The distributed schemes are performed with neighbouring information [12-17]. Zhang *et al.* [18] introduced a distributed gradient algorithm to realize optimal economical generation control. Further, Zhang *et al.* [19] selected the incremental costs of each DGs unit as a consensus variable to minimize the total operation cost in a distributed manner. And Yang *et al.* [20] presented another consensus algorithm for performing the equal incremental costs of each DG with strongly connected communication topology to solve the economical dispatch problem. However, [18-20] are highly dependent on communications for information exchange, and the central controller is unnecessary compared to the centralized schemes.

In contrast, some scholars proposed the decentralized approaches to deal with the power dispatch problem, where require no communications [21-23]. Droop control method as decentralized approach has been widely used in microgrids [24-25]. By emulating the behaviour of a synchronous generator, this well-known control technique aims to proportionally share active and reactive power with adjusting frequency and output voltage amplitudes of each inverter locally [26-27]. However, the power sharing cannot guarantee the economy of microgrid in most cases. In order to reduce the total active generation costs (TAGC) of microgrids via decentralized approach, Nutkani *et al.* [7] presented the linear droop schemes by introducing maximum or mean generation costs to the droop coefficient, where the lower-cost DG holds higher priority of output power. Actually, the generation cost of DGs is a nonlinear function of active output power, thus the TAGC of microgrids might not be optimized efficiently. Further, applying the nonlinear cost functions of DGs to typical droop scheme, Nutkani *et al.* [28] proposed a nonlinear cost-based droop control scheme. But the optimal economical operation of system is not obtained yet. Cingoz *et al.* [29] proposed a nonlinear droop control strategy based on polynomial fitting method to realize economical operation of microgrids.

Although the economical operation could be achieved with the methods mentioned above, they are not the optimal economical operation approaches in most cases. Moreover, these works are focus on the microgrid with paralleled inverters. Nowadays, the microgrid with cascaded inverters has been recognized as an important alternative in the medium voltage market for microgrid applications [30]. The cascaded inverters fed by DGs in microgrids are capable of achieving high-quality output voltages and input currents with low harmonic content [31-32]. However, there are few studies about the economic dispatch problem for this structure especially via decentralized approach.

To address these concerns, this paper proposes an optimal decentralized economical-sharing scheme for the microgrid with cascaded inverters. It applies the frequency and current as carriers to implement the economical dispatch among DGs without communications. The proposed scheme is carried off-line according to the optimal power sharing function of the total loads, and is the incapability of plug-and-play. The proposed scheme is fully decentralized and requires no communication network among DGs, therefore, it offers increased reliability. Finally, the effectiveness of proposed scheme has been verified through simulations and experiments.

The rest of the paper is organized as follows. Section 2 shows the problem formulation based on the total active generation costs and the power demand of the mricrogrid. The power transmission characteristics of the microgrid with cascaded inverters are discussed in Section 3. And the proposed optimal economical-sharing scheme is introduced in Section 4. A stability analysis of the proposed scheme is presented in Section 5. Then, the simulation validations in Section 6 and the experimental results in Section 7 are provided to verify the effectiveness and performance of the proposed scheme. Finally the paper is concluded in Section 8.

## 2. Problem formulation

Assume that there are $n$ diapatchable DGs in an islanded microgrid. With the aim of minimizing the total active generation cost, the economical dispatch problem is formulated as:

$$\begin{aligned} & \min\left(\sum C_i(P_i)\right) \\ s.t. \quad & \sum P_i = P_L \end{aligned} \quad (1)$$

where $P_i, P_L, C_i(P_i)$ are the output active power, load requirements, cost function of $i^{th}$ DG, respectively, and $i \in \{1,2,\cdots,n\}$. The feasible ranges of (1) are on a closed interval $[P_{i,\min}, P_{i,\max}]$. For different load $P_L$, there is the corresponding optimal solution $\left(P_1^*, P_2^*, \cdots, P_n^*\right)$ for the economical operation of microgrids. Note that the optimal dispatch $P_i^*$ could be regarded as a map of $P_L$:

$$P_i^* = \xi_i(P_L) \quad (2)$$

where $\xi_i(P_L)$ is a function of $P_L$. Without lose generality, $\xi_i(P_L)$ is continuously on $[P_{L,\min}, P_{L,\max}]$. The function $\xi_i(P_L)$ could be achieved off-line:

- If the function $C_i(P_i)$ is simple, such as the positive definite quadratic functions [28], $\xi_i(P_L)$ can be calculated by analytical method.
- If the function $C_i(P_i)$ is complicated, $\xi_i(P_L)$ could be obtained by fitting method, intelligent algorithm, and so on.

## 3. Power transmission characteristics of microgrids with cascaded inverters

The considered microgrid comprises a series of cascaded inverters [30-32] shown in Fig. 1 (a), and its equivalent circuit is shown in Fig. 1 (b). The obtained load voltage is presented as:

$$V_x e^{j\delta_x} = z' z_x \sum_{i=1}^{n} V_i e^{j\delta_i} \quad (3)$$

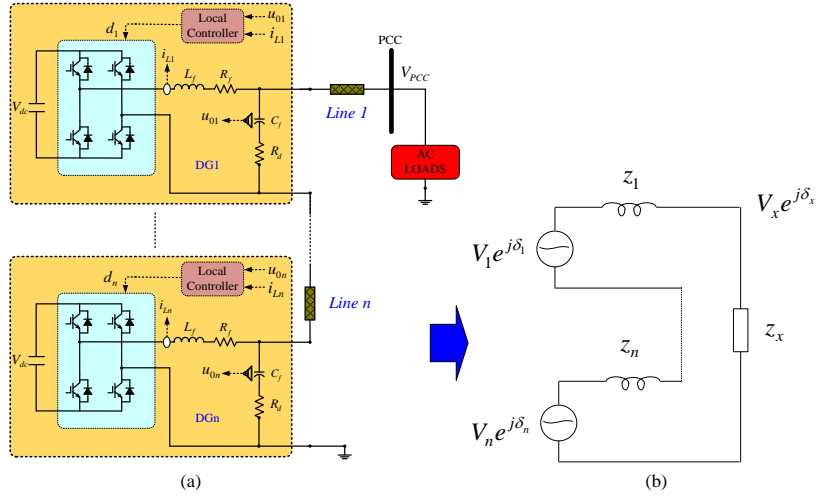

**Fig. 1.** *(a) Considered microgrids with cascaded inverters (b) equivalent circuit.*

$$z' = \frac{1}{z_x + \sum_{i=1}^{n} z_i} \tag{4}$$

where $z_x$, $z_i$, and $z'$ are load impedance, line impedance, and equivalent impedance, respectively. And $z'$ could be rewritten as:

$$z' = Z'e^{j\theta'} \tag{5}$$

where $\theta'$ is the phase of $z'$. After obtaining the load voltage, it is easy to get the expressions of output active power $P_i$ and reactive power $Q_i$ of $i^{th}$ DG:

$$P_i = V_i |Z'| \sum_{j=1}^{n} V_j \cos(\delta_i - \delta_j - \theta') \tag{6}$$

$$Q_i = V_i |Z'| \sum_{j=1}^{n} V_j \sin(\delta_i - \delta_j - \theta') \tag{7}$$

Accordingly, it is concluded that the power transmission characteristics of this structure could be summarized as follows:

- $P_i$ and $Q_i$ of each DG could be regulated by changing the phase difference between $\delta_i$ and $\delta_j$.
- Each DG shares a common current in the microgrid.

## 4. Proposed decentralized economical-sharing scheme

Assume $P_i > 0$, $Q_i > 0$, the proposed optimal decentralized economical-sharing scheme for AC microgrids with cascaded inverters is:

$$\begin{cases} f_i = f_{\min} + \dfrac{h}{\varphi_i} P_i \\ V_i = \varphi_i V_{PCC} \end{cases} \tag{8}$$

$$\varphi_i = \dfrac{g_i(I)}{\sum_{i=1}^{n} g_i(I)} \tag{9}$$

where $h$, $I$, $V_{PCC}$ are positive constant, current, voltage at PCC, respectively, and $g_i(I)$ is a function of $I$ for $i^{th}$ DG.

When all DGs get into steady state,

$$f_1 = f_2 = \cdots = f_n \tag{10}$$

$$P_1 : P_2 : \cdots : P_n = g_1(I) : g_2(I) : \cdots : g_n(I) \tag{11}$$

By the characteristics of cascaded inverters, there yields:

$$P_1 : P_2 : \cdots : P_n = V_1 : V_2 : \cdots : V_n \tag{12}$$

$$Q_1 : Q_2 : \cdots : Q_n = V_1 : V_2 : \cdots : V_n \tag{13}$$

Because the voltage of the PCC is regarded as a constant, for different load $P_L$, there is a corresponding current $I$ of microgrid. And $P_L$ can be deemed to a map of $I$, (2) could be written as follows:

$$P_i^* = g_i(I) \tag{14}$$

Neglecting the voltage loss of line, then

$$V_1 + V_2 + \cdots + V_n = V_{PCC} \tag{15}$$

Combining (12), (14)-(15), yields:

$$V_i = \dfrac{g_i(I)}{\sum_{i=1}^{n} g_i(I)} V_{PCC} \tag{16}$$

The proposed scheme can realize the optimal economical operation for the microgrid with cascaded inverters under the communications unnecessary, control the frequency within the allowable ranges, and maintain the voltage at PCC in rated conditions. The global variables, frequency $f$ and current $I$, are applied to achieve the optimal economical operation only with the local information.

## 5. Stability analysis

To investigate the stability of the microgrid under the proposed scheme, the small-signal analysis method [33-35] is applied. Without loss of generality, the microgrid shown in Fig. 1 is studied. Neglecting the dynamics of fast-time scale, the dynamic equations of $i^{th}$ DG are written as following:

$$\begin{cases} \dot{\delta}_i = \omega_{min} + \dfrac{h}{\varphi_i} P_{if} \\ V_i = \varphi_i V_{PCC} \end{cases} \quad (17)$$

where $\omega_{min} = 2\pi f_{min}$. The Filtered active power $P_{if}$ and reactive power $Q_{if}$ could be written as:

$$P_{if} = P_i \dfrac{w_c}{s + w_c} \Rightarrow \dot{P}_{if} = (P_i - P_{if}) w_c \quad (18)$$

$$Q_{if} = Q_i \dfrac{w_c}{s + w_c} \Rightarrow \dot{Q}_{if} = (Q_i - Q_{if}) w_c \quad (19)$$

where $w_c$ is the filter angular frequency. Assume that $\omega_s$ is the synchronous frequency in steady state. Let $\delta_s = \int \omega_s dt$, and denote $\tilde{\delta}_i = \delta_i - \delta_s$, then the frequency in (17) is rewritten as:

$$\dot{\tilde{\delta}}_i = \omega_{min} - \omega_s + \dfrac{h}{\varphi_i} P_{if} \quad (20)$$

The small-signal model of (18-20) around the operating point is given as follows:

$$\Delta \dot{P}_{if} = w_c \sum_{j=1}^{n} \dfrac{\partial P_i}{\partial \tilde{\delta}_j} \Delta \bar{\delta}_j - \Delta P_{if} w_c \quad (21)$$

$$\Delta \dot{Q}_{if} = w_c \sum_{j=1}^{n} \dfrac{\partial Q_i}{\partial \tilde{\delta}_j} \Delta \bar{\delta}_j - \Delta Q_{if} w_c \quad (22)$$

$$\Delta \dot{\tilde{\delta}}_i = \dfrac{h}{\varphi_i} \Delta P_{if} \quad (23)$$

Express (21-23) in the form of matrix:

$$\dot{\mathbf{X}} = \mathbf{A}\mathbf{X} \quad (24)$$

$$\mathbf{X} = \begin{bmatrix} \Delta P_{1f} & \cdots & \Delta P_{nf} & \Delta Q_{1f} & \cdots & \Delta Q_{nf} & \Delta \tilde{\delta}_1 & \cdots & \Delta \tilde{\delta}_n \end{bmatrix}^T \quad (25)$$

$$A = \begin{bmatrix} -w_c \mathbf{I}_{n \times n} & \mathbf{0}_{n \times n} & \mathbf{T}_{P\delta} \\ \mathbf{0}_{n \times n} & -w_c \mathbf{I}_{n \times n} & \mathbf{T}_{Q\delta} \\ \mathbf{T}_\delta & \mathbf{0}_{n \times n} & \mathbf{0}_{n \times n} \end{bmatrix} \quad (26)$$

where

$$T_{P\delta} = \begin{bmatrix} \dfrac{\partial P_1}{\partial \tilde{\delta}_1} & \cdots & \dfrac{\partial P_1}{\partial \tilde{\delta}_n} \\ \vdots & \ddots & \vdots \\ \dfrac{\partial P_n}{\partial \tilde{\delta}_1} & \cdots & \dfrac{\partial P_n}{\partial \tilde{\delta}_n} \end{bmatrix}; \quad T_{Q\delta} = \begin{bmatrix} \dfrac{\partial Q_1}{\partial \tilde{\delta}_1} & \cdots & \dfrac{\partial Q_1}{\partial \tilde{\delta}_n} \\ \vdots & \ddots & \vdots \\ \dfrac{\partial Q_n}{\partial \tilde{\delta}_1} & \cdots & \dfrac{\partial Q_n}{\partial \tilde{\delta}_n} \end{bmatrix}; \quad I = diag[1 \; \cdots \; 1]_{n\times n}; \quad T_\varphi = diag[\dfrac{1}{\tilde{\varphi}_1} \; \cdots \; \dfrac{1}{\tilde{\varphi}_n}]_{n\times n};$$

To test the stability of the proposed scheme, the root-locus method is applied. The studied microgrid comprises three DGs, and the parameters are same as simulation validations in Section 6. The root locus is depicted by changing the loads and the filter angular frequency $w_c$. There is a simple eigenvalue at zero corresponding to rotational symmetry in Fig. 2 and Fig. 3, which has been proven in [36]. Fig. 2 shows the root locus as the load resistance decreases from 18Ω to 6Ω, in which all eigenvalues are in the left half-plane. Thus the considered microgrid is stable under the load increasing. With $w_c$ changing from $80\pi$ to $120\pi$ rad/s, the poles are in the left half-plane in Fig.3. Therefore, the stable operation of the microgrid could be achieved in terms of the proposed scheme.

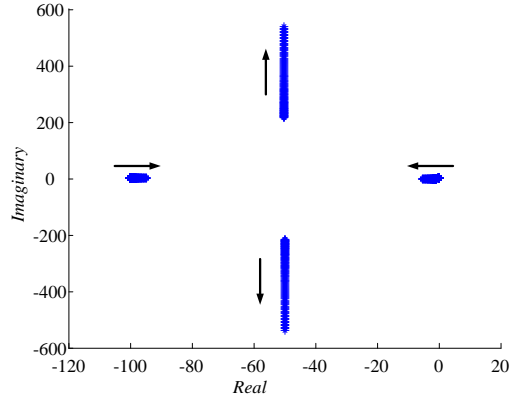

***Fig. 2.*** *Stable root with slower dynamics as the load increases ( $w_c = 100\pi$ rad/s).*

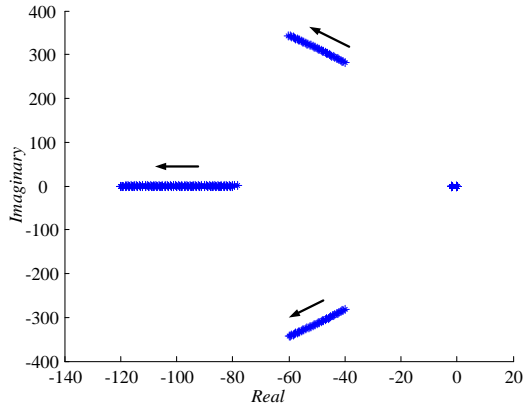

***Fig. 3.*** *Root locus by increasing $w_c$ (the load resistance is 12 Ω).*

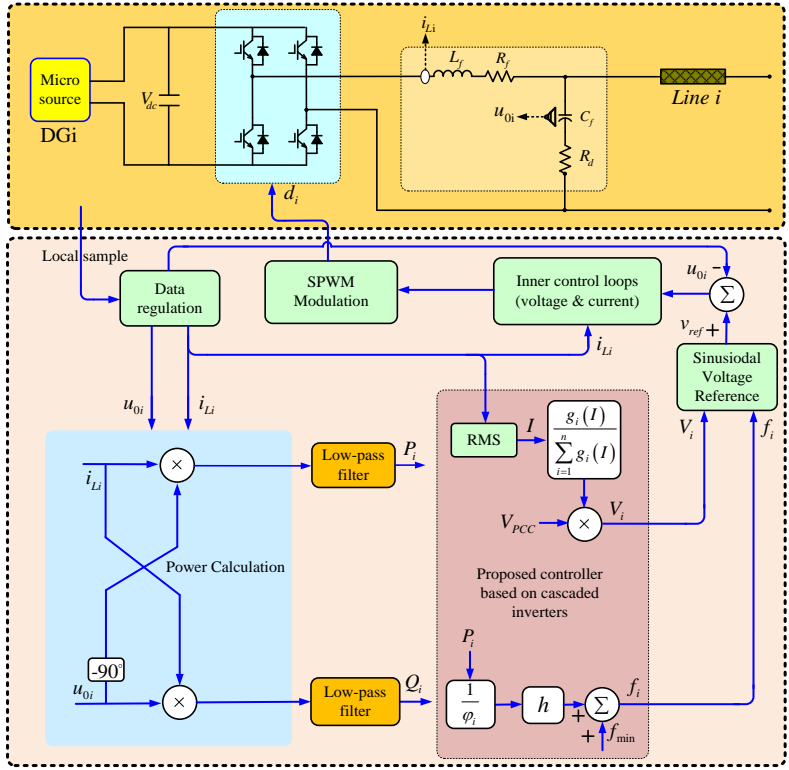

***Fig. 4.*** *Configuration of the single-phase DG unit.*

**Table 1** Parameters for simulations (see Fig. 1)

| Parameters | Values | Parameters | Values |
|---|---|---|---|
| $f$ (Hz) | [49, 51] | $L_{Line1}$ (H) | 1.5e-3 |
| $V_{PCC}$ (V) | 110 | $L_{Line2}$ (H) | 1.6e-3 |
| $L_f$ (H) | 1.5e-3 | $L_{Line3}$ (H) | 1.2e-3 |
| $R_f$ ($\Omega$) | 0.4 | $P_{\max}$ (W) | 1000 |
| $C_f$ ($\mu$F) | 20 | $Q_{\max}$ (Var) | 1000 |
| $R_d$ ($\Omega$) | 3.3 | $h$ | 0.1 |

## 6. Simulation validations

The proposed economical-sharing scheme is verified through MATLAB/Simulink. The considered microgrid with three DGs is employed (see Fig. 1). And the associated parameters are listed in Table 1. And the generation cost of $i^{th}$ DG is obtained from [28]: $C_1(P_1)=0.25P_1^2$, $C_2(P_2)=0.15P_2^2$, $C_3(P_3)=0.1P_3^2+0.01P_3$. From (2), the optimal solution $P_i^*$ is calculated as the map of $P_L$, $P_1^*=\xi_1(P_L)=6P_L/31+3/310$, $P_2^*=\xi_2(P_L)=10P_L/31+1/62$, $P_3^*=\xi_3(P_L)=15P_L/31-4/155$. From (14), $P_L$ is the map of $I$, then the optimal solution could be rewritten as: $P_1^*=g_1(I)=660I/31+3/310$, $P_2^* = g_2(I) = 1100I/31 + 1/62$, $P_3^* = g_3(I) = 1650I/31 - 4/155$. According to the proposed economical sharing scheme: $V_1 = 660/31 + 3/(310I)$, $V_2 = 1100 + 3/(1 I)$, $V_3 = 1650/31 - 4/(155I)$, and $I > 0$. The detailed configuration of the single-phase DG unite is depicted in Fig. 4.

### 6.1. Case1: Performance of the proposed scheme as load changes

In this case, the DGs are regulated in terms of the proposed scheme, and the fluctuations of the total active power loads are shown in Fig. 5(a). The load demands are scheduled as 0.683*p.u*, 1.35*p.u*, 2*p.u* in the interval [0s, 1s], [1s, 2s], [2s, 3s], respectively. From Fig. 5(b), the frequency of all DGs reaches a common value quickly regardless of load demands variations. It is illustrated that the proposed scheme could adjust all DGs synchronous operation, and controls the frequency within the allowable ranges. When the microgrid gets into steady state, the frequency would converge to a certain value, and yields $\frac{h}{\varphi_i} P_i$ equal as load changes shown in Fig. 5 (c). And the corresponding active power sharing among DGs is shown in Fig. 5 (d). Accordingly, the proposed scheme could regulate all DGs in microgrid synchronous operation, and realize the optimal economical active power sharing among DGs in the presence of load changes.

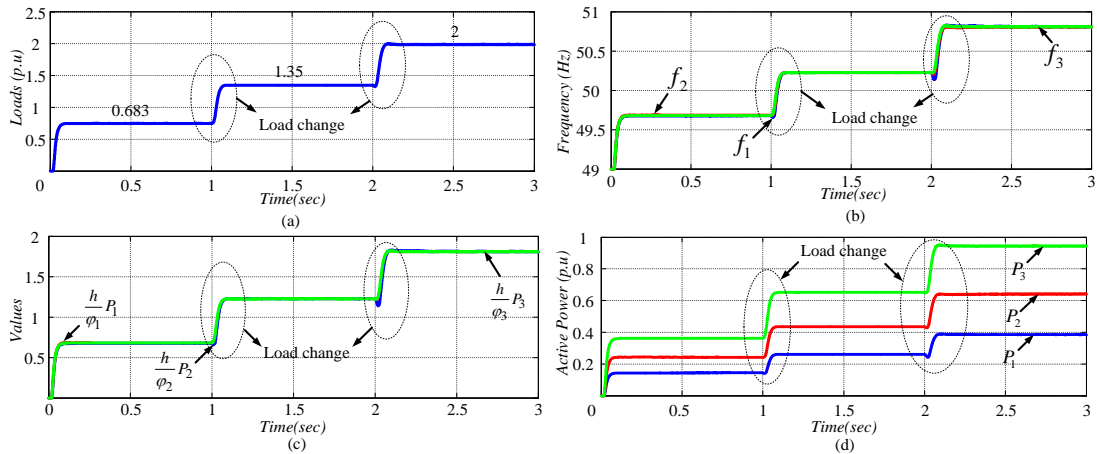

***Fig. 5.*** *Variations of (a) the total active power demands (b) frequency over time (c) $\frac{h}{\varphi_i} P_i$ (d) optimal active power sharing as load changes.*

The reactive power load demands shown in Fig. 6 (a) are 0.185*p.u*, 0.366*p.u*, 0.541*p.u* in the interval [0s, 1s], [1s, 2s], [2s, 3s], respectively. And the reactive power sharing among DGs with load increasing is shown in Fig. 6 (b). The corresponding ratio of the reactive power relative to $Q_1$ is shown in Fig. 6 (c). Similarly, the sharing ratio of the active power with the proposed scheme is shown in Fig. 6 (d).

The voltage variations at PCC are shown in Fig. 6, in which it is around 110V as the load changes. Therefore, the allocation ratio of the active and reactive power among DGs is equal over time. The frequency and voltage at PCC could be controlled within the allowable ranges.

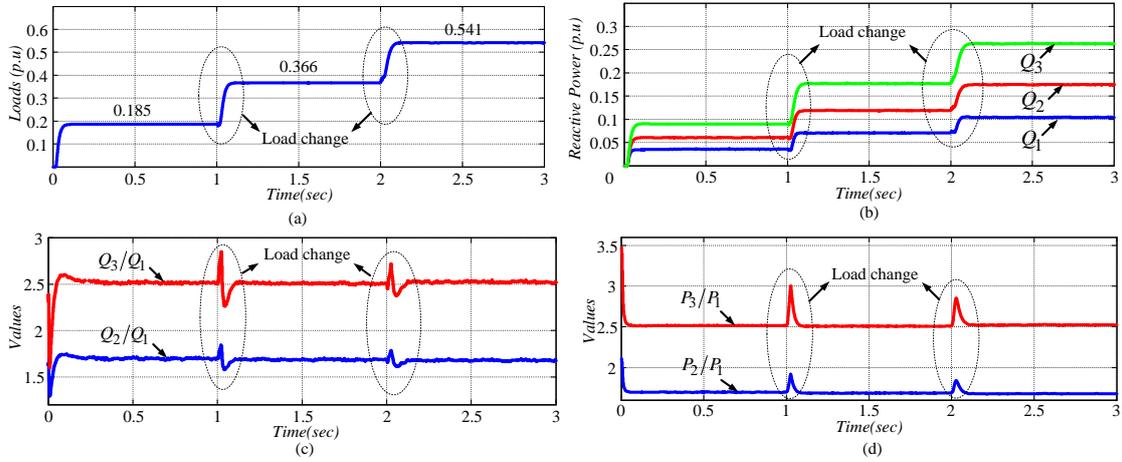

*Fig. 6.* Variations of (a) the reactive power demands (b) the reactive power sharing (c) the ratio of reactive power (d) the ratio of active power over time.

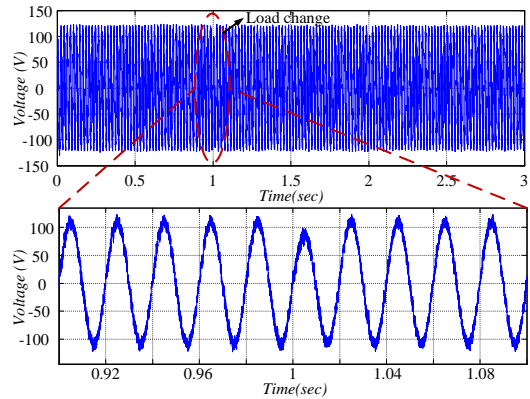

*Fig. 7.* Simulation waveforms of the voltage at PCC with load stepping.

### 6.2. Case2: Economy comparisons for the proposed scheme and proportional dispatch scheme

In this case, the simulations are implemented with comparison the TAGC for the proportional sharing scheme and the proposed scheme with the same load changes shown in Fig. 5. The variations of the TAGC are calculated in terms of the proportional-sharing scheme shown in Fig. 8 (a). Under the same setting, the proposed scheme is carried out, and the corresponding TAGC is calculated in Fig. 8 (b). Based on the simulation results, the proposed scheme always obtains the lowest TAGC as the load changes. Therefore, it is concluded that the proposed scheme is a low-cost solution.

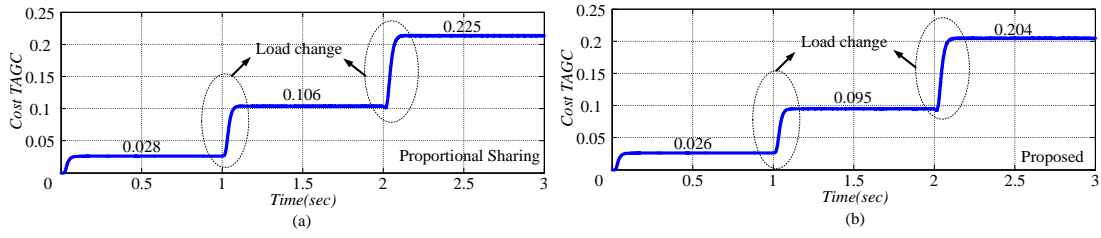

*Fig.8. Variations of TAGC with (a) proportional sharing scheme (b) proposed scheme.*

## 7. Experimental results

A microgrid prototype shown in Fig. 9 is built in lab to verify the effectiveness and performance of the proposed method. It comprises two DGs based on the single phase voltage source inverters which are controlled by digital signal processors (TMS320f28335) and the sampling rate is 12.8 kHz. DG2 and DG3 are considered and the corresponding generation characteristics are same as the simulation validations in Section 6. The experimental parameters are shown in Table 2.

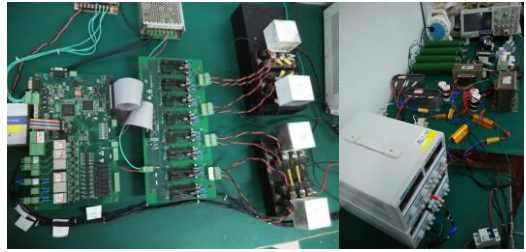

*Fig.9. Prototype setup of the microgrid with cascaded inverters.*

**Table 2** Parameters for experiments (see Fig. 1)

| Parameters | Values | Parameters | Values |
|---|---|---|---|
| $f$ (Hz) | [49, 51] Hz | $R_d\ (\Omega)$ | 5 |
| $V_{PCC}$ (V) | 100V | $L_{Line2}$ (H) | 0.3e-3 |
| $L_f$ (H) | 0.6e-3 | $L_{Line3}$ (H) | 0.6e-3 |
| $R_f\ (\Omega)$ | 0.5 | $P_{max}$ (W) | 200 |
| $C_f\ (\mu F)$ | 20 | $Q_{max}$ (Var) | 100 |

### 7.1. Case1: Synchronous operation with load changes

Due to the limitations of experimental conditions, the microgrid comprises only two DGs. In this case, synchronous operation of DGs in microgrid with cascaded inverters has been verified as the proposed scheme implemented. And the experimental waveform is shown in Fig. 10. From the experimental result, the voltage of DG2 and DG3 and the current of the microgrid are same frequency and phase over time. This characteristics are retained even load changes. Therefore, the proposed scheme could satisfy the requirement of DGs' synchronous operation and maintain the microgrid stable operation.

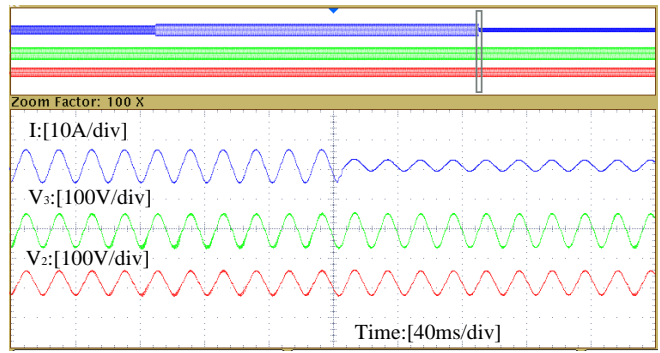

***Fig.10.*** *Experimental waveforms with load changes.*

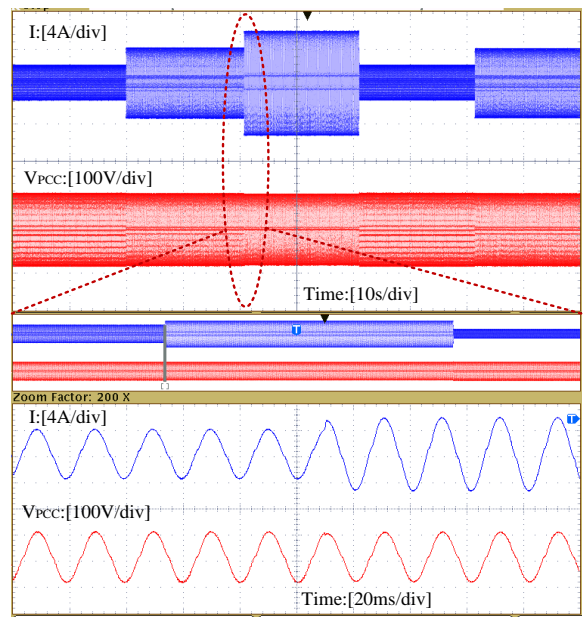

***Fig.11.*** *Experimental voltage and current waveforms with proposed scheme.*

### 7.2. Case2: Performance of the proposed scheme as load changes

In this case, the experiment is implemented in terms of the proposed scheme. The experimental voltage and current waveforms at PCC under load variations are shown in Fig. 11. The active power loads over time are shown in Fig.12 (a). When the microgrid is in steady state, the frequency converges to a certain point shown in Fig. 12 (b). And the proposed scheme could drive DGs synchronous operation and maintain the microgrid stability. The active power allocations of DG2 and DG3 with the proposed scheme are illustrated in Fig. 12 (c). The reactive power sharing among DGs with the same ratio of active power allocations are shown in Fig. 12 (d).

Based on the experimental result, the proposed scheme could drive DGs synchronization and realize the microgrid stable operation, the frequency and voltage could be controlled within the feasible ranges.

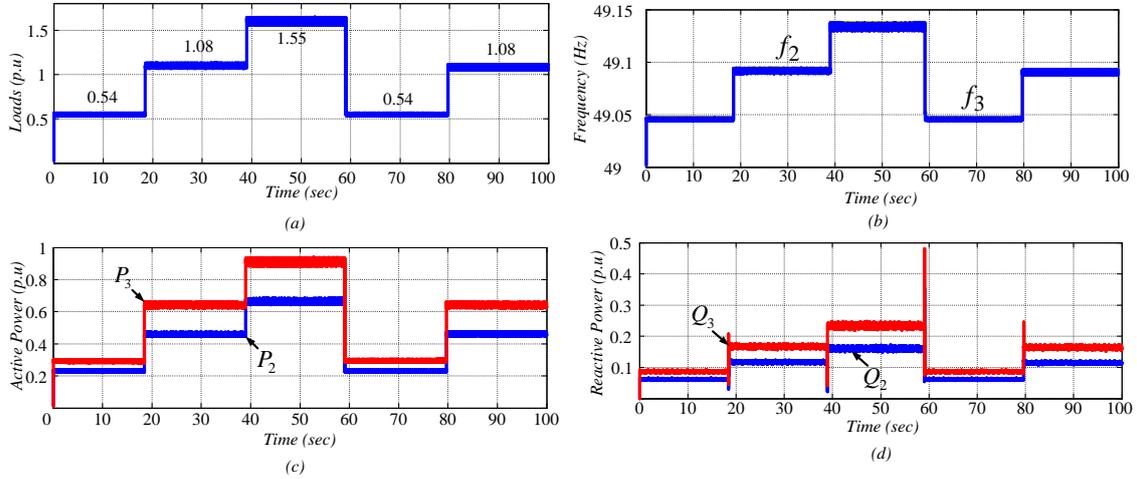

***Fig.12.*** *Variations of (a) active power load demands (b) frequency (c) active power sharing (d) reactive power sharing.*

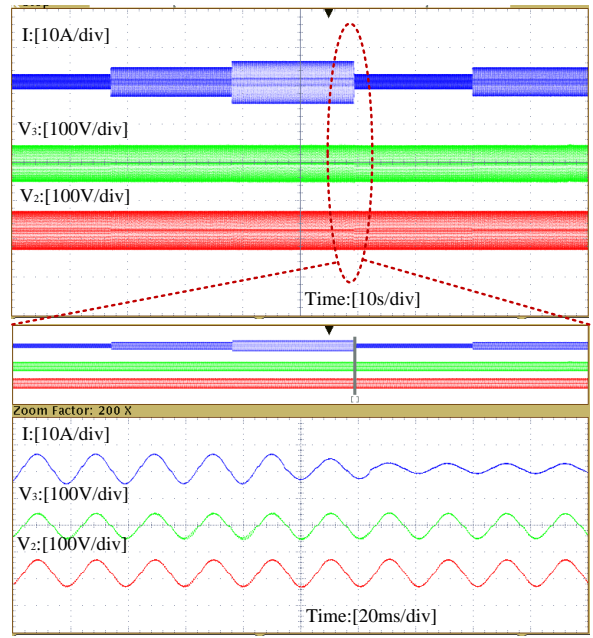

***Fig.13.*** *Experimental waveforms with load changes.*

### 7.3. Case3: Economy comparisons for the proposed scheme and proportional dispatch scheme

In this case, the economy is compared with the proposed scheme and proportional dispatch scheme under the same setup. The experimental voltage and current waveforms are shown in Fig. 13. And the active power demands of the microgrid are shown in Fig. 12 (a). Under the implementation of proportional dispatch scheme, the frequency variations over time are shown in Fig. 14 (a). And the corresponding active power sharing among DGs is shown in Fig. 14 (b). The TAGC is calculated with the proportional dispatch scheme shown in Fig. 14 (c). Similarly, the TAGC with the proposed scheme is shown in Fig. 14 (d), which is always lower than Fig. 14 (c).

Accordingly, the proposed scheme could realize the optimal economical sharing while maintaining the smooth and stable operation even in presence of load stepping.

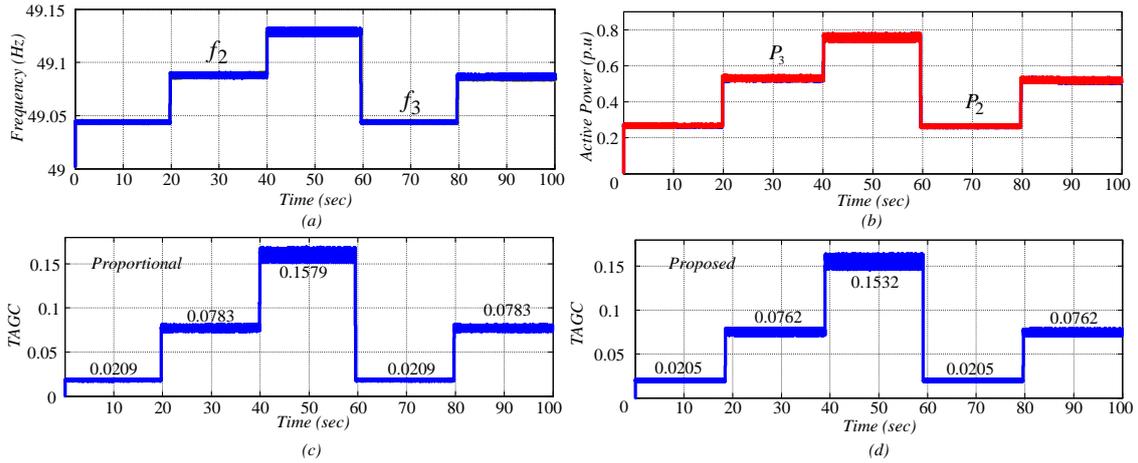

***Fig.14.*** *Variations of (a) frequency (b) active power sharing (c) TAGC with proportional dispatch scheme (d) TAGC with proposed scheme.*

## 8. Conclusion

In this paper, an optimal decentralized economical-sharing scheme is proposed to reduce the total active generation costs to a minimum for microgrids with cascaded inverters. The optimal power delivery is implemented through the global variables frequency and current. Because the implementation of the proposed method only needs the local information of each DG, communications are not needed. Therefore, it is a reliable and low-cost solution. Moreover, the frequency and voltage at PCC could be controlled within the allowable ranges. The construction of the proposed scheme is carried off-line, and the simplicity is preserved and hence more likely to meet industry requirement. The simulations and experiments have been verified the effectiveness and performance of the proposed method.

The aim of this work is to show the validity of the proposed optimal decentralized economical-sharing scheme. And the more work would be covered in future, such as integrating DGs power ratings, realizing four-quadrant operation.

## 9. References

[1] Z. Zhang, X. Huang, and J. Jiang, "A load-sharing control scheme for a microgrid with a fixed frequency inverter," *Electric Power Systems Research*, vol. 80, pp. 311–317, doi:10.1016/j.epsr.2009.09.014, May. 2011.

[2] R. J. Hamidi, H. Livani, and S.H. Hosseinian, "Distributed cooperative control system for smart microgrids," *Electric Power Systems Research*, vol. 130, pp. 241–250, doi:10.1016/j.epsr.2015.09.012, Sep. 2015.

[3] W. Fei, J. Duarte, Hendrix, M.A.M, "Grid-interfacing converter systems with enhanced voltage quality for microgrid application concept and implementation," *IEEE Trans. Power Electron.*, vol. 26, no. 12, pp. 3501–3513, Apr. 2011.

[4] R. Moslemi, J. Mohammadpour, "Accurate reactive power control of autonomous microgridsusing an adaptive virtual inductance loop," *Electric Power Systems Research*, vol. 129, pp. 142–149, doi:10.1016/j.epsr.2015.08.001, Aug. 2015.

[5] P. Arboleya, D. Diaz, and J.M. Guerrero, "An improved control scheme based in droop characteristic for microgrid converters," *Electric Power Systems Research*, vol. 80, pp. 1215–122, doi:10.1016/j.epsr.2010.04.003, Jun. 2010.

[6] H. Han, Y. Liu, and Y. Sun, "An improved droop control strategy for reactive power sharing in islanded microgrid," *IEEE Trans. Power Electron.*, vol. 30, no. 6, pp. 3133–3141, Jun. 2015.

[7] I. U. Nutkani, P. C. Loh, and P. Wang, "Linear decentralized power sharing schemes for economic operation of AC microgrids," *IEEE Trans. Industrial Electronics*, vol. 63, no. 1, pp. 225-234, Jan. 2016.

[8] N. Nikmehr, S. N. Ravadanegh, "Optimal Power Dispatch of Multi-Microgrids at Future Smart Distribution Grids," *IEEE Trans. Smart Grid*, vol. 6, no. 4, pp.1648–1657, Jul.2015.

[9] E. Barklund, N. Pogaku, M. Prodanovic, C. Hernandez-Aramburo, and T. C. Green, "Energy management in autonomous microgrid using stability-constrained droop control of inverters," *IEEE Trans. Power Electron.*, vol. 23, no. 5, pp. 2346–2352, Sep. 2008.

[10] A. G. Tsikalakis and N. D. Hatziargyriou, "Centralized control for optimizing microgrids operation," *IEEE Trans. Energy Convers.*, vol. 23, no. 1, pp. 241–248, Mar. 2008.

[11] F. Katiraei, R. Iravani, N. Hatziargyriou, and A. Dimeas, "Microgrids management," *IEEE Power Energy Mag*, Vol. 6, no. 3, pp. 54–65, May. 2008.

[12] Yao, S., Chaolu, Z., Xiaochao, H., *et al.*: 'Distributed cooperative synchronization strategy for multi-bus microgrids', International Journal of Electrical Power and Energy Systems, vol. 86, pp. 18-28, Mar. 2017.

[13] M. Yazdanian, and A. Mehrizi-Sani, "Distributed control techniques in microgrids,"*IEEE Trans. Smart Grid,* vol. 5, no. 6, pp. 2901–2909, 2014.

[14] Q. Shafiee, J. M. Guerrero, and J. C. Vasquez, "Distributed secondary control for islanded microgrids—A novel approach," *IEEE Trans. Power Electron.*, vol. 29, no. 2, pp. 1018–1031, 2014.

[15] J. W. Simpson-Porco, Q. Shafiee, F. D¨orfler, J. C. Vasquez, J. M.Guerrero, and F. Bullo, "Secondary frequency and voltage control of islanded microgrids via distributed averaging," *IEEE Trans. Ind.Electron.*, vol. 62, no. 11, pp. 7025–7038, 2015.

[16] H. Xin, Z. Lu, Y. Liu, and D. Gan, "A center-free control strategy for the coordination of multiple photovoltaic generators," *IEEE Trans. Smart Grid*, vol. 5, no. 3, pp. 1262–1269, 2014.

[17] V. Nasirian, Q. Shafiee, J. M. Guerrero, F. L. Lewis, and A. Davoudi,"Droop-free distributed control for AC microgrids," *IEEE Trans. Power Electron.*, vol. 31, no. 2, pp. 1600–1617, 2016.

[18] W. Zhang, W. Liu, X. Wang, L. Liu, and F. Ferrese, "Online optimalgeneration control based on constrained distributed gradient algorithm,"*IEEE Trans. Power Syst.*, vol. 30, no. 1, pp. 35–45, 2015.

[19] Z. Zhang and M.-Y. Chow, "Convergence analysis of the incremental cost consensus algorithm under different communication network topologies in a smart grid," *IEEE Trans. Power Syst.*, vol. 27, no. 4, pp. 1761–1768, 2012.

[20] S. Yang, S. Tan, and J.-X. Xu, "Consensus based approach for economic dispatch problem in a smart grid," *IEEE Trans. Power Syst.*, vol. 28,no. 4, pp. 4416–4426, 2013.

[21] J. M. Guerrero, H. Lijun, and J. Uceda, "Control of distributed uninterruptible power supply systems," *IEEE Trans. Ind. Electron.*, vol. 55, no. 8, pp. 2845–2859, Jul. 2008.

[22] J. M. Guerrero, M. Chandorkar, T.-L. Lee, and P. C. Loch, "Advanced control architectures for intelligent microgrids—Part I: Decentralized and hierarchical control," *IEEE Trans. Ind. Electron.*, vol. 60, no. 4, pp.1254–1262, 2013.

[23] D. E. Olivares, A. Mehrizi-Sani, A.H. Etemadi, "Trends in microgrid control," *IEEE Trans. Smart Grid*, vol. 5, no. 4, pp. 1905–1919, 2014.

[24] N. Jaleeli, L. S. VanSlyck, D. N. Ewart, L. H. Fink, and A. G. Hoffmann,"Understanding automatic generation control," *IEEE Trans. Power Syst.*,vol. 7, no. 3, pp. 1106–1122, Aug. 1992.

[25] Yajuan Guan, Josep M. Guerrero, and Xin Zhao,"A New Way of Controlling Parallel-Connected Inverters by Using Synchronous-Reference-Frame Virtual Impedance Loop—Part I: Control Principle," *IEEE Trans. Power Electron.*, VOL. 31, NO. 6, pp. 4576-4593, Jun. 2016.

[26] P. Piagi and R. H. Lasseter, "Autonomous control of microgrids," *IEEE Power Engineering Society General Meet.*, *Montreal, Canada*, 2006. DOI: 10.1109/PES.2006.1708993.

[27] C. T. Lee, "A new droop control method for the autonomous operation of distributed energy resource interface converters," *IEEE Trans. Power Electron.*, vol. 28, no. 4, pp. 1980–1993, Apr. 2013.

[28] I. U. Nutkani, P. C. Loh, and F. Blaabjerg, "Droop scheme with consideration of operating costs," *IEEE Trans. Power Electron.*, vol. 29, no. 3, pp. 1047-1052, May. 2014.

[29] F. Cingoz, A. Elrayyah, Y. Sozer, "Plug and play nonlinear droop construction scheme to optimize microgrid operations," *Energy Conversion Congress and Exposition (ECCE), Pittsburgh, PA,* Sept. 2014, pp. 76 – 83.

[30] A. Mortezaei, M. G. Simões, and A. S. Bubshait, "Multifunctional Control Strategy for Asymmetrical Cascaded H-Bridge Inverter in Microgrid Applications," *IEEE Trans. Industry Applications*, early access, DOI: 10.1109/TIA.2016.2627521.

[31] M. Malinowski, K. Gopakumar, J. Rodriguez, and M. A. Perez, "A survey on cascaded multilevel inverters," *IEEE Trans. Ind. Electron.*, vol. 57, no. 7, pp. 2197–2206, Jul. 2010.

[32] R. R. Karasani, V. B. Borghate, and P. M. Meshram, "A Three-Phase Hybrid Cascaded Modular Multilevel Inverter for Renewable Energy Environment," *IEEE Trans. Power Electronics*, vol. 32, no. 2, pp. 1070–1087, Feb. 2017.

[33] N. Pogaku, M. Prodanovic and T. Green, "Modeling, analysis and testing of autonomous operation of an inverter-based microgrid," *IEEE Trans. Power Electron*, vol. 22, no. 2, pp. 613–625, Mar. 2007.

[34] Y. A.-R. I. Mohamed and E. F. El-Saadany, "Adaptive decentralized droop controller to preserve power sharing stability of paralleled inverters in distributed generation microgrids," *IEEE Trans. Power Electron*, vol. 23, no. 6, pp.2806–2816, Dec. 2008.

[35] X. Q. Guo, Z. G. Lu, and B. C. Wang, "Dynamic Phasors-Based Modeling and Stability Analysis of Droop-Controlled Inverters for Microgrid Applications," *IEEE Trans. Smart Grid*, vol. 5, no. 6, pp. 2980-2987, Nov. 2014.

[36] J. W. Simpson-Porco, F. Dörfler, and F. Bullo, "Synchronization and power sharing for droop-controlled inverters in islanded microgrids," *Automatica,* vol. 49, no. 9, pp. 2603–2611, Sep. 2013.